\documentclass[12pt]{JHEP3}
\usepackage{amsmath,amstext,amsbsy,amssymb}
\newcommand{\Bb}{\mathbf{B}}
\newcommand{\Eb}{\mathbf{E}}
\newcommand{\Db}{\mathbf{D}}

\newcommand{\fr}{\frac}
\newcommand{\lb}{\label}
\newcommand{\ti}{\tilde}
\newcommand{\be}{\begin{equation}}
\newcommand{\ee}{\end{equation}}
\newcommand{\ba}{\begin{array}}
\newcommand{\ea}{\end{array}}
\newcommand{\beqa}{\begin{eqnarray}}
\newcommand{\al}{\alpha}
\newcommand{\bt}{\beta}
\newcommand{\ro}{\rho}

\newcommand{\La}{\Lambda}
\newcommand{\De}{\Delta}
\newcommand{\si}{\sigma}
\newcommand{\te}{\theta}

\newcommand{\del}{\partial}
\newcommand{\eeqa}{\end{eqnarray}}
\newcommand{\ep}{\epsilon}

\newcommand{\kd}{\delta}

\newcommand{\mbf}{\mathbf}

\preprint{hep-th/0407269}

\title{Equivalence of partition functions for noncommutative 
$\mathbf{ U(1)}$
gauge theory and its dual in phase space}

\author{\"{O}mer F. DAYI \\
 Physics Department, Faculty of Science and
Letters, Istanbul Technical University,
 80626 Maslak--Istanbul,
Turkey\\
 Feza G\"{u}rsey Institute,
 P.O.Box 6, 81220
\c{C}engelk\"{o}y--Istanbul, Turkey \\
E-mail: \email{ dayi@itu.edu.tr,  dayi@gursey.gov.tr} }

\author{Bar\i\c{s} YAPI\c{S}KAN  \\
 Physics Department, Faculty of Science and
Letters, Istanbul Technical University,
 80626 Maslak--Istanbul,
Turkey\\
E-mail: \email{ yapiska1@itu.edu.tr.} }

\abstract{Equivalence of partition functions for
$U(1)$ gauge theory and its dual in 
appropriate phase spaces is established 
in terms of constrained hamiltonian formalism of
their parent action. Relations between the
electric--magnetic duality transformation
and the (S) duality transformation
which inverts the  strong coupling domains to
the weak coupling domains
of noncommutative $U(1)$ gauge theory are discussed
in terms of the
lagrangian and the hamiltonian densities.
The approach presented for the commutative
case is utilized to demonstrate that
noncommutative $U(1)$ gauge theory and its dual
possess the same partition function in their phase spaces
at the first order in the noncommutativity
parameter $\theta.$ }


\begin{document}

\section{Introduction}

Maxwell equations in vacuum are 
electric--magnetic duality invariant. 
Similarly one can  formulate  a duality transformation
of $U(1)$ gauge theory action:
This (S) duality inverts   weak coupling constant regions
into strong coupling constant regions.
A parent action \cite{bus} is defined  in terms of
the dual 
gauge field $A_D$ and the antisymmetric second rank tensor F. When 
it is employed in the path integral, if one integrates over $A_D$ 
the partition function of the ordinary $U(1)$ theory results. Instead 
of $A_D$ one can integrate F which yields the partition function of
the dual $U(1)$ theory. Thus, one can easily show
 equivalence of  partition functions for
the $U(1)$ and its dual theory,  up to a normalization 
constant. On the other hand hamiltonian
description of these theories are shown to be connected by a canonical
transformation and as a consequence it followed that the
partition functions in their phase spaces are the same\cite{loz}.
We demonstrate that this equivalence  
can directly be obtained in terms of  hamiltonian formulation
of the  parent action: Utilizing constraints one can integrate the desired 
phase space variables obtaining either the  partition function
of $U(1)$ gauge theory or the partition function of
its dual theory in appropriate 
phase spaces.

In terms of fields taking values in a noncommutative space
one can introduce
a noncommutative $U(1)$ gauge theory. 
However, these noncommuting fields 
can be mapped into ordinary fields
utilizing the  Seiberg--Witten map\cite{sw}. 
Then, a dual noncommutative $U(1)$ action can be
obtained analogous to the commutative case, by introducing a parent 
action\cite{grs}. When the initial $U(1)$ theory possesses a spatial
noncommutativity the dual one is also  noncommutative  
$U(1)$ gauge theory 
whose
time coordinate is noncommuting with spatial coordinates.
Hamiltonian formulation of the latter theory, which
is suitable to study the noncommutative D3--brane, was
presented in \cite{dy}. 

Although electric--magnetic duality transformation
is an invariance of Maxwell equations in vacuum, it is
known that it maps the lagrangian density to itself up
to an overall minus sign and keeps intact the hamiltonian
density of $U(1)$ gauge theory. 
Electric--magnetic duality 
transformation of the equations of motion
of  noncommutative $U(1)$ theory 
is studied in \cite{abt}. In spite of that,
we would like to
understand the relation between electric--magnetic 
duality and the (S) duality inverting strong and weak
 coupling regimes. Hence, we discuss relations
of the electric--magnetic duality with
the dual description of  the noncommutative gauge theory utilizing
the lagrangian and  the hamiltonian densities.
We only deal with the first order 
approximation in the noncommutativity
parameter $\theta.$

For $U(1)$ gauge theory the parent action can be 
used in the related path integrals  to derive  
the duality symmetry between 
the original and the dual theories. But, for the noncommutative theory
one should employ the  equations of motion derived from the parent action,
to obtain the  dual noncommutative $U(1)$ theory action\cite{grs}.
The (S) duality symmetry  of the 
noncommutative $U(1)$ theory was not established  and relation
between their partition functions was unknown. 
We show that partition functions for the  noncommutative
$U(1)$ theory with spatial noncommutativity and its
dual whose time coordinate is effectively noncommuting
with spatial coordinates, are equivalent in the appropriate phase
spaces. To achieve this we follow the approach
presented for the commutative gauge theory.

In Section 2 we first present the
constrained Hamiltonian structure 
of the  parent action of  Maxwell theory. 
The partition function
of the parent action 
in phase  space is written. We show that 
by integrating over the appropriate fields either the
partition  function of $U(1)$ theory in its phase space
or the partition function of its dual theory in
dual  phase space results.

Relations between 
the electric--magnetic duality transformations and the (S) dual
actions of noncommutative
$U(1)$ gauge theory are discussed  in terms of configuration
space fields as well as in terms of  phase space fields in
Section 3. 

In Section 4, guided by  the approach of Section 2,
the path integral of parent action for the   noncommutative theory
in phase space is studied.
We demonstrate  equivalence of partition functions for  spatially 
noncommutative
$U(1)$ gauge theory and its dual being effectively space--time 
noncommutative  $U(1)$ gauge theory
with an inverted coupling constant, at the first order in
$\te .$

\section{Partition functions for
$U(1)$ gauge theory and its dual}

In  Minkowski space--time with   the 
metric $g_{\mu \nu }= {\rm diag} (-1,1,1,1),$
$U(1)$ gauge theory and its dual 
can be extracted from the parent action 
\be
\lb{spi}
S_P=-\int d^4 x \left(\fr{1}{4g^2}F_{\mu\nu} F^{\mu\nu}+\fr{1}{2}
\ep_{\mu\nu\rho\sigma} 
\del^\mu A_D^\nu  F^{\rho\sigma}\right).
\ee
Here, 
$F_{\mu\nu}$ are not related to  gauge fields, 
they are the basic variable fields. 
Let us introduce the canonical 
momenta $P_{\mu\nu}$ and $P_{D\mu}$ corresponding
to $F_{\mu\nu}$ and  $A_{D\mu}$. Definitions of
the canonical momenta $P_D^\mu,\ P_{\mu \nu} ,$
yield  the weakly vanishing 
primary constraints 
\beqa
\Phi^1_{\mu\nu} \equiv  P_{\mu\nu}\approx 0, \lb{p1} \\
\xi^1 \equiv P_{D0}\approx 0 \lb{x1}, \\
\chi^2_i \equiv 
P_{Di}+\frac{1}{2}\ep_{ijk}F^{jk} \approx 0.\lb{p2}
\eeqa
Canonical hamiltonian associated with the parent action
 (\ref{spi}) is
\be
\lb{hpc}
H_{PC}=\int 
d^3x\ \left[ \fr{1}{2g^2}F^{0i}F_{0i}+\fr{1}{4g^2}F^{ij}F_{ij}-\fr{1}{2}
\ep_{ijk}\del^i{{A_D}^0}F^{jk}+\ep_{ijk}\del^i{A_D}^jF^{0k}
\right] .
\ee
Consistency of the primary constraints (\ref{p1})--(\ref{p2}) 
with the equations of motion resulting from (\ref{hpc})
gives
rise to 
the secondary constraints
\beqa
\Phi^3 
\equiv \{P_{D0},H_{PC} \} =
\ep_{ijk}\del^iF^{jk}\approx 0,  \lb{p3}\\
\chi^4_i 
\equiv \{ P_{0i},H_{PC} \}  =
F_{0i}+  g^2 \ep_{ijk}\del^j{A_D}^k 
\approx 0 .\lb{p4}
\eeqa

Let us find out the number of physical phase space fields:
The constraint  (\ref{x1}) 
is obviously first class.
Besides it, the
linear combination
\be
\lb{x2}
\xi^2\equiv \del_i \chi^2_i -\fr{1}{2}\Phi^3=
\del_iP_{Di} \approx 0,
\ee
is also a  first class constraint. 
A vector can be completely described by giving its 
divergence and rotation (up to a boundary condition). (\ref{x2}) is 
derived taking 
divergence of $\chi_i^2,$ so that, there are still
two linearly independent second class constraints following
from the curl of $\chi_i^2.$
Obviously, the constraints $\Phi^1,\ \Phi^3,\ \chi^4 $ 
are all second class and linearly independent.
Therefore, the number of  physical phase space fields is four.

To deal with path integrals, we choose the
gauge fixing (subsidiary) conditions
\be
\La^1=A_{D0}\approx 0,\ \La^2=\del_i A_{Di}\approx 0
\ee
for the first class constraints (\ref{x1}) 
and (\ref{x2}). The linearly independent second class
constraints resulting from the curl of $\chi^2_i$
can be taken as
\be
\lb{sccs}
\Phi^2_n \equiv 
 C_n^i\chi_i^2 \equiv K_n^i\ep_{ijk}\del_j\chi_k^2 \approx 0, 
\ee
where $n=1,2,$ and $K_n^i$ are some constants which should be chosen 
in accordance with  solutions of the other constraints when
they vanish strongly. Instead of dealing with 
$\chi^4_i$ we introduce another set of linearly
independent second class constraints:
\be
\lb{scc2}
\Phi^4_n \equiv 
 M_n^i\chi_i^4 \equiv L_n^i\ep_{ijk}\del_j\chi_k^4\approx 0,\ \
\Phi^4_3 \equiv \del^iF_{0i} \approx 0 . 
\ee
$L_n^i$ are some constants. As we will see, explicit forms of
$K_n^i$ and 
$L_n^i$ play no role in our calculations.

Partition function associated with the 
hamiltonian (\ref{hpc}) in the total phase space is
\be
\lb{partpa}
Z=\int DA_D DF DP_D DP\  
\De  
\exp \left\{ i \int d^3x\left[ P_{D\mu} \dot{A}_D^{\mu} 
+ P_{\mu\nu}\dot{F}^{\mu\nu}
- H_{PC}\right] \right\}.
\ee
We suppressed the indices of the integration
variables and the measure $\De$ is 
defined\cite{fra},\cite{sen} as
\be
\De = 
\det \{\xi^\al,\La^\beta \}
{\rm det}^{1/2} \{\Phi^a,\Phi^b\} 
\prod_{\si =1}^2\kd(\xi^\si)\kd(\La^\si)
\prod_{c =1}^{4}   \kd(\Phi^c ) .
\ee
The determinant related to first class constraints and
their subsidiary conditions 
is 
\[
\det \{\xi^\al,\La^\beta \}= \det \del_i\del^i
\equiv \det (\del^2 ).
\]
The determinant due to the second  class constraints 
can be calculated as
\be
\lb{dsc}
{\rm det}^{1/2} \{\Phi^a,\Phi^b\}
= \det \left( \ep_{ijk}\del^iC_1^jC_2^k  \right) 
\det \left( \ep_{ijk}\del^iM_1^jM_2^k  \right),
\ee
where the linear differential operators $C_n^i$ and $M_n^i$ are defined in 
(\ref{sccs}) and (\ref{scc2}). Here, the determinants of these 
linear operators should be interpreted as multiplication of
their eigenvalues. 

Performing functional integrations over the variables
$F^{\mu \nu }$,$P_{\mu\nu}$ and $A_D^0,P_D^0$ 
we obtain 
\beqa
Z & = &\int 
D\mbf{A}_D D\mbf{P}_D 
\kd(\mbf{\del} \cdot \mbf{P}_D)
\kd(\mbf{\del} \cdot \mbf{A}_D) 
{\rm det}(\del^2)
\nonumber \\
&& \exp \left\{ i \int d^3x\left[P_{Di}\dot{A}_D^i-
\fr{1}{2g^2}P_{Di}P_D^i
-\fr{g^2}{4}F_D^{ij}F_{Dij}\right] \right\}.   \lb{zhd}
\eeqa
Here, the factor 
${\rm det}^{1/2} \{\Phi^a,\Phi^b\} $
is canceled 
with the determinant arising from the Dirac delta functions
$\kd (\Phi^a)$  when we use them 
to express $F_{\mu\nu},\ P_{\mu\nu}$ in terms of the ``physical"
fields $\mbf{A}_D,\ \mbf{P}_D.$ Although here this can be
observed by direct calculation\footnote{To obtain (\ref{zhd})
we do not need to deal with the set (\ref{scc2}). It is easier 
to  employ  (\ref{p4}) with an appropriate determinant.},
it is true in general 
when one gets rid
of second class constraints by imposing them strongly
and deal with  reduced phase space  path integrals\cite{sen}.

Now, in  (\ref{partpa})  we  would like to
perform integrations  over the 
dual fields $A_{D\mu},\ P_{D\mu}$ and the momenta $P_{\mu \nu}.$
Vanishing of the constraint (\ref{p3}) 
strongly, i.e. $\Phi^3=0,$ dictates that 
\be
\lb{far}
F_{ij}=\del_iA_j-\del_jA_i.
\ee
Being a second class constraint $\Phi^3=0$
should  eliminate one phase space  variable. 
However, the
number of  independent components of $F_{ij}$ and $A_i$
are the same. So that, solving  $\Phi^3=0$
as (\ref{far}) and
dealing with $A_i$ instead of $F_{ij},$ has to
be accompanied with a 
condition on $A_i.$
The constraint (\ref{p2}) involves 
only curl of $A_i,$ therefore,  $\Phi^2_n=0$ 
give information only about the two components 
of $A_i.$
In order to describe
$A_i$ completely one needs to furnish its divergence.
Thus, we choose as the missing condition 
\be
\lb{scon}
\del_iA^i=0.
\ee

After performing 
$ A_{D\mu},P_{D\mu}$ and
$P_{\mu\nu}$ 
integrations 
in (\ref{partpa}) we obtain
\beqa
Z&=& \det g^{-4}\int D\mbf{A} DF_{0j}
{\rm det}(\del^2)
\kd(\del^l F_{0l}) \kd(\mbf{\del} \cdot \mbf{A}) \nonumber \\
& &  \exp \left\{i\int d^3x[-\fr{1}{g^2}F_{0i}\dot{A}^i
+ \fr{1}{2g^2}F^{0i}F_{0i}- \fr{1}{4g^2}F^{ij}F_{ij}]\right\}. \lb{zbo}
\eeqa
We used the fact that  expressing
$A_{Di}$ and $P_{Di}$ in terms of the ``physical''
fields $A_i,\ F_{0i},$
using  
the Dirac delta functions $\kd (\Phi^a) 
\kd(\mbf{\del} \cdot \mbf{P}_D)
\kd(\mbf{\del} \cdot \mbf{A}_D) ,$  contributes
to the measure as
\[
\left[ \det g^4 \det (\del^2)
 \det \left( \ep_{ijk}\del^iC_1^jC_2^k  \right) 
\det \left( \ep_{ijk}\del^iM_1^jM_2^k  \right) \right]^{-1}.
\]
Moreover,
here $F_{ij}$ is given by (\ref{far})
and we performed the  change of variables
$F_{ij} \to A_i.$ 
We choose  domains
of the integrals such that in (\ref{partpa})  
we can perform the replacement
\be
\lb{nor}
DF_{ij} \kd (\ep^{klm}\del_kF_{lm})
\kd (C_n^i( P_{Di}+\frac{1}{2}\ep_{ijk}F^{jk}) ) 
 \to \det (\del^2) 
DA_i \kd (\del_jA^j)
\kd \left(C_n^i( P_{Di}+\ep_{ijk}\del^jA^k) \right). 
\ee
One can observe that $\det (\del^2)$ should be included in the
measure when one deals with the gauge fields $A_i$  satisfying the
condition (\ref{scon}), considering this change of variables
from the beginning with an appropriate change of the momenta
$P_{ij}\to P_i$ where the latter are canonical momenta
of $A_i.$ 
  
Observe that in (\ref{zbo})
the variables $F_{0i}$ 
 can be renamed as 
\be
\lb{f0p}
F_{0i}=-g^2 P_i,
\ee
where $P_i$ are the canonical momenta associated to  $A_i.$
Thus, (\ref{zbo}) becomes
\beqa
Z& = & \det g^{-4} \int D\mbf{A} D\mbf{P}
{\rm det} (\del^2) 
\kd(\mbf{\del} \cdot \mbf{P})
\kd(\mbf{\del} \cdot \mbf{A})
\nonumber \\
&&\exp \left\{ i\int d^3x[P_i\dot{A}^i-\fr{g^2}{2}P_i P^i
-\fr{1}{4g^2}F^{ij}F_{ij}]\right\} \lb{zoo2}.
\eeqa

Although  (\ref{f0p}) is resulted after performing
functional integrals in (\ref{partpa}), we could derive
it from the constraint structure using  the Dirac brackets:
\beqa
\{F_{0i}(x),P_{Dj}(y)\}_{\rm Dirac} & = & 
\{F_{0i},P_{0k}\}
\{P_{0k},\Phi^4_l\}^{-1}
\{\Phi^4_l,P_{Dj}\} \nonumber \\
& & = g^2\ep_{ikj}\fr{\del \kd^3 (x-y)}{\del x_k}. \lb{db}
\eeqa
Making use of (\ref{far}) in $\chi^2_i=0$ yields
\be
\lb{pea}
P_{Di}=-\ep_{ijk}\del^jA^k .
\ee
Plugging (\ref{pea}) into the left hand side
of  the Dirac bracket (\ref{db}), leads to
\be
-\ep_{jkl}\fr{\del}{\del y_k}
\{F_{0i}(x),A_{l}(y)\}_{\rm Dirac}  = 
g^2\ep_{ijk} \fr{\del \kd^3 (x-y)}{\del x_k},
\ee
which is solved as
\be
\{F_{0i}(x),A_{j}(y) \}_{\rm Dirac}  =g^2 \kd_{ij}\kd^3 (x-y) .
\ee
Thus, (\ref{f0p}) follows.

We choose the normalization such that  partition function
for Maxwell theory in hamiltonian formalism is given by
\beqa
Z_H\equiv Z_N(g)& = & \det g^{-2} \int D\mbf{A} D\mbf{P}
{\rm det} (\del^2) 
\kd(\mbf{\del} \cdot \mbf{P})
\kd(\mbf{\del} \cdot \mbf{A})
\nonumber \\
&&\exp \left\{ i\int d^3x[P_i\dot{A}^i-\fr{g^2}{2}P_i P^i
-\fr{1}{4g^2}F^{ij}F_{ij}]\right\} \lb{ytmh}.
\eeqa
We denoted the normalized partition function as $Z_N(g)$.
The normalized partition function of the dual theory
in phase space is
\beqa
Z_{HD} = Z_N(g^{-1}) & = &
\det g^2 \int 
D\mbf{A} D\mbf{P}{\rm det}(\del^2)
\kd(\mbf{\del} \cdot \mbf{P})
\kd(\mbf{\del} \cdot \mbf{A}) 
\nonumber \\
&& \exp \left\{ i \int d^3x\left[P_{i}\dot{A}^i-
\fr{1}{2g^2}P_{i}P^i
-\fr{g^2}{4}F^{ij}F_{ij}\right] \right\},   \lb{ythd}
\eeqa
where we renamed $A_D^i,\ P_D^i$ as $A^i,\ P^i .$
By comparing Z obtained in (\ref{zhd}) and (\ref{zoo2}) we
conclude that in hamiltonian formalism partition functions for
Maxwell theory and its dual are the same
\be
Z_H =Z_{HD},
\ee
which can equivalently  be written in terms of the normalized 
partition functions as

\be
\lb{peq}
Z_N (g) =Z_N (g^{-1}).
\ee
This result was obtained in \cite{loz} in terms of  canonical
transformations without gauge fixing factor  and with 
another normalization.

\section{Relations between the electric-magnetic duality 
and  the dual actions of
noncommutative $U(1)$ theory }

Noncommuting coordinates are operators even at the classical 
level. In spite of this  fact we can treat them as the usual commuting
coordinates  by replacing 
operator products with 
$*$--products. Utilizing the latter   a noncommutative $U(1)$
gauge theory is defined which can be written in terms of the usual gauge 
fields,
after performing the Seiberg--Witten map, as\cite{sw} 
\be
\lb{sncu}
S_{NC}=-\fr{1}{4g^2} \int d^4x \left( F_{\mu\nu}F^{\mu\nu}+
2\te^{\mu\nu}F_{\nu\ro}F^{\ro\si}F_{\si\mu}-
\fr{1}{2}\te^{\mu\nu}F_{\mu\nu}F_{\ro\si}F^{\ro\si}\right),
\ee
at the first order in the noncommutativity 
parameter $\te^{\mu\nu}$ which is  constant and antisymmetric. 
Here $F_{\mu\nu}= \del_\mu A_\nu - \del_\nu A_\mu .$
Dual of (\ref{sncu}) is obtained in \cite{grs} as
\be
\lb{sncud}
S_{NCD}=-\fr{g^2}{4} \int d^4x\left( F_D^{\mu\nu}F_{D\mu\nu}
+2\ti{\te}^{\mu\nu}F_{D\nu\rho}F_D^{\rho\si}F_{D\si\mu}
-\fr{1}{2}\ti{\te}^{\mu\nu}F_{D\mu\nu}F_{D\rho\si}F_D^{\rho\si}
\right),
\ee
where $F_{D\mu\nu}= \del_\mu A_{D\nu}
- \del_\nu A_{D\mu}$ and 
\[
\ti{\te}^{\mu\nu} = \fr{g^2}{2} \ep^{\mu \nu \rho \si}\te_{\rho \si }.
\]
Obviously, if the original theory (\ref{sncu}) possesses spatially
noncommutative coordinates, in the dual theory (\ref{sncud})
time is effectively noncommuting with spatial coordinates.

We would like to discover relations between  electric--magnetic 
duality and the (S) duality transformation for noncommutative $U(1)$
gauge theory in configuration space. Let us write the actions 
(\ref{sncu}) and (\ref{sncud}) in terms of the electric
and magnetic fields:
When the magnetic field vector
\be
\lb{mag} 
B_i=-\fr{1}{2} \ep_{ijk}F^{jk}
\ee
and the electric field vector $E_i=F_{0i}$
are employed, the original 
action (\ref{sncu}) becomes\cite{kru}
\be
\lb{act1}
S_{NC}=\int d^4x \left[\fr{1}{2g^2}(\Eb^2-\Bb^2)(1-\mathbf{\te} \cdot \Bb )
+\fr{1}{g^2}\mathbf{\te} \cdot \Eb \Eb \cdot \Bb  \right] ,
\ee
where  the vector $\mathbf{\te}$ is defined by 
$\te^{ij}=\ep^{ijk}\te_k.$

For the dual case we adopt the same notation:
 $E_i=F_{D0i}$ and
\be
\lb{magd}
B_i=-\fr{1}{2} \ep_{ijk}F_D^{jk}.
\ee
Hence, the dual action (\ref{sncud}) can be written as
\be
\lb{act2}
S_{NCD}=\int d^4x \left[ \fr{g^2}{2}(\Eb ^2-\Bb ^2)(1+\ti{\mathbf{\te}}
\cdot \Eb )
-g^2\ti{\mathbf{\te}}\cdot \Bb \Eb \cdot \Bb \right],
\ee
where $\ti{\mathbf{\te}}$ vector is 
defined as $\ti{\te}^i\equiv \ti{\te}^{0i}.$

One can observe that under the transformation
\be
\lb{fra}
\Eb \to g^2\Bb , \quad
 \Bb\to -g^2\Eb ,
\ee
(\ref{act1}) is mapped
into  the dual action (\ref{act2}) up to  an overall  minus sign.
This is a well known property of abelian gauge theory
action. Thus, it  persists in the noncommutative theory.

We also would like to 
obtain relations between the
electric--magnetic duality and
the (S) duality transformations of the noncommutative $U(1)$ theory
in hamiltonian formalism. Canonical hamiltonian
associated with
(\ref{sncu}) can  be derived as
\beqa
H_{NC}&=& \int d^3x{\Big [}\fr{g^2}{2}P_i^2+ \fr{1}{4g^2}F_{ij}F^{ij}
+\fr{1}{2g^2}\te^{ij}F_{jk}F^{kl}F_{li}-\fr{1}{8g^2}\te^{ij}
F_{ij}F_{kl}F^{kl} \nonumber \\
&& + g^2\te^{ij}P_jP^kF_{ki}-\fr{g^2}{4}\te^{ij}F_{ji}P_k^2{\Big ]}, \lb{ncho}
\eeqa
where we choose the subsidiary condition $A_0=0$ which corresponds
to the constraint $P_0=0.$ Furthermore, there 
is the constraint 
$\del_i P^i=0$.  
Hamiltonian of the dual  noncommutative $U(1)$ gauge theory (\ref{sncud})
is obtained in \cite{dy} by two different approaches as\footnote{There are some 
misprints
in \cite{dy} which are corrected here.} 
\beqa
H_{NCD}=\int && d^3x{\Big [}\fr{1}{2g^2}P^2_{Di}+ \fr{g^2}{4}F_{Dij}F^{ij}_D +
\fr{1}{2g^4}\ti{\te}_{0i}P^i_D P^2_{Dj} + \fr{1}{4}\ti{\te}_{0i}P^i_D
F_{Djk}F^{jk}_D \nonumber \\ 
&& + \ti{\te}_{0i}F^{ij}_D F_{Djk}P^k_D{\Big ]} \lb{nchd}
\eeqa
with the constraint $\del_i P^i_D=0$ after setting
$P_{D0}=0,\ A_{D0}=0. $

Let us  introduce 
the vector  field
$P_i =g^{-2}D_i$
and the magnetic fields as before (\ref{mag}). Hence, we 
write the hamiltonian (\ref{ncho}) as
\be
\lb{nca}
H_{NC}=\int d^3x \left[ \fr{1}{2g^2}(\Db^2+\Bb^2)-\fr{1}{2g^2}
\mathbf{\te} \cdot \Bb (\Bb^2-\Db^2)-\fr{1}{g^2}\mathbf{\te} \cdot \Db 
\Bb \cdot \Db \right].
\ee
Similarly, let us introduce  $P_{Di}=g^2 D_i$
and the magnetic field as in (\ref{magd}).
Then, the hamiltonian (\ref{nchd}) becomes
\be
\lb{hncd}
H_{NCD}=\int d^3x \left[ \fr{g^2}{2}(\Db^2+\Bb^2)- \fr{g^2}{2}
\ti{\mathbf{\te}}\cdot \Db (\Db^2-\Bb^2)- 
g^2 \ti{\mathbf{\te}}\cdot \Bb \Bb \cdot \Db \right] .
\ee
One can show that 
under the map
\be
\lb{tra}
\Db\to -g^2 \Bb, \quad \Bb \to g^2 \Db
\ee
the hamiltonian (\ref{nca}) transforms into 
 the dual hamiltonian  (\ref{hncd}).
Thus, noncommutative electric-magnetic
duality transformation in  hamiltonian formulation
is given by (\ref{tra}).
Observe that  the
lagrangian 
and the hamiltonian description  of electric--magnetic duality
transformations, (\ref{fra}) and (\ref{tra}), 
seem to be ``inverted". 

Definition of the canonical momenta $P_i$
following from (\ref{sncu}) can be used to express  
$P_i$
 in terms of
the electric field $E_i=F_{0i}.$
Then, one  can express the  hamiltonian (\ref{ncho}) as\cite{kru} 
\be
\lb{hlmc}
H_{NC}=\int d^3x 
\left[ \fr{1}{2g^2}(\Eb^2+\Bb^2)(1-\mathbf{\te}\cdot \Bb)+\fr{1}{g^2}
\mathbf{\te} \cdot \Eb \Eb \cdot \Bb \right].
\ee
Analogously, 
the canonical momenta $P_{Di}$
derived from
 (\ref{sncud}) can be  expressed   in terms of
the electric field $E_i=F_{D0i}.$ Making use of it
in the  hamiltonian (\ref{nchd}) one obtains
\be
\lb{hlmd}
H_{NCD}=\int d^3x \left[ \fr{g^2}{2}(\Eb^2+\Bb^2)+ g^2 
\ti{\mathbf{\te}}\cdot \Eb \Eb^2 \right].
\ee
(\ref{hlmc}) and
(\ref{hlmd}) are not related with a transformation resembling
the electric--magnetic duality transformation (\ref{fra}).

Electric--magnetic duality transformation 
of the noncommutative hamiltonians cannot be given
in terms of $\Eb ,\ \Bb$ fields but using $\Db,\ \Bb .$
This is an expected result: Hamiltonians should be
written in the momenta $P_i$ or $P_{Di}$ not by using
the ``velocities" $F_{0i}$ or $F_{D0i}.$ In the commuting
case this difference does not appear due to the fact that
$\mathbf{P}=\Eb .$

\section{Partition functions for the noncommutative $U(1)$ theory
and its dual}

The noncommutative  $U(1)$ action 
(\ref{sncu}) and its dual (\ref{sncud})
can be derived from the parent action\cite{grs} 
\be
S_{NP}  = - \fr{1}{4g^2} \int d^4x {\Big (} 
F_{\mu\nu}F^{\mu\nu}+2\te^{\mu\nu}
F_{\nu\ro}F^{\ro\si}F_{\si\mu}- \fr{1}{2}\te^{\mu\nu}F_{\nu\mu}
F_{\ro\si}F^{\si\ro} 
+ \fr{1}{2}  \ep_{\mu\nu\ro\si}A_D^\mu\del^\nu F^{\ro\si} {\Big )}, 
\lb{npar}
\ee
where $F_{\mu\nu}$ are not composed of any other field. 
We only deal with the first order approximation in $\te_{\mu\nu}.$ 
To acquire  hamiltonian formalism we introduce the canonical momenta
$P_{\mu \nu},\ P_{D\mu}$ corresponding to 
the configuration space variables
$F_{\mu \nu},\ A_{D\mu}.$ Definitions of the canonical momenta
$P_D^\mu$ and $P_{\mu\nu}$ yield the
primary constraints 
\beqa
\ti{\Phi}^1_{\mu\nu} \equiv P_{\mu\nu}\approx 0, \lb{np1}\\
\ti{\xi}^1 \equiv P_{D0}\approx 0 , \lb{nx1}\\
\ti{\chi}^2_i \equiv P_{Di}+\fr{1}{2}\ep_{ijk}F_{jk}\approx 0,  \lb{np2}
\eeqa
which are weakly vanishing. One can show that 
canonical hamiltonian related to (\ref{npar}) is
\beqa
H_{NPC} & = & \int d^3x{\Big [}-\fr{1}{2}\ep_{ijk}\del^i A_D^0F^{jk}
+\ep_{ijk}\del^i A_D^j F^{0k} +\fr{1}{2g^2}F_{0i}F^{0i} \nonumber \\
&&+\fr{1}{4g^2}F_{ij}F^{ij}+\fr{1}{g^2}F^{0i}F_{ij}\te^{jk}F_{k0}
+\fr{1}{2g^2}F^{ij}F_{jk}\te^{kl}F_{li} \nonumber \\
&&-\fr{1}{4g^2}\te^{ij}F_{ij}F_{0k}F^{0k} -\fr{1}{8g^2}\te^{ij}F_{ij}
F_{kl}F^{kl} {\Big ]}. \lb{nch}
\eeqa
Consistency of the primary constraints (\ref{np1})--(\ref{np2})
with the hamiltonian equations of motion originating from (\ref{nch}),
leads to the
secondary constraints 
\be
  \ti{\Phi}^3
\equiv \{P_{D0},H_{NPC} \} = 
 \ep_{ijk}\del^iF^{jk}\approx 0 ,  \lb{zo2} 
\ee
\beqa
 \ti{\chi}^4_i  \equiv \{ P_{0i},H_{NPC} \} & =& 
F^{0i}-F_{ij}\te^{jk}F_{k0}
-F^{0j}F_{jk}\te^{ki} -\fr{1}{2}\te^{jk}F_{kj}F_{0i}  \nonumber \\
& & -g^2\ep_{ijk}\del^jA_D^k \approx 0 .   \lb{np4}   
\eeqa
Like the commuting case $\ti{\xi}^1$  and  the
linear combination of (\ref{np2}) and (\ref{zo2}) 
\be
\lb{sla}
\ti{\xi}^2 \equiv \del_i \ti{\chi}^2_i - \fr{1}{2} \ti{\Phi}^3
= \del_i P_D^i \approx 0 
\ee
are first class constraints. 
Curl  of $\chi^2_i$ leads to  
two linearly independent second class
constraints:
\be
\lb{sccsn}
\ti{\Phi}^2_n \equiv 
 C_n^i\ti{\chi}_i^2 \equiv K_n^i\ep_{ijk}\del_j
\ti{\chi}_k^2 \approx 0, 
\ee
where $n=1,2.$ 
Analogous to 
the commuting case, instead of $\ti{\chi}^4_i$  we deal with 
the following set of second class constraints
\beqa
 & \ti{\Phi}^4_n \equiv 
 M_n^i\ti{\chi}_i^4  \equiv L_n^i\ep_{ijk}\del_j\ti{\chi}_k^4\approx 0,
&   \lb{scc2n1} \\
 & \ti{\Phi}^4_3 \equiv \del_i\left(
F^{0i}-F_{ij}\te^{jk}F_{k0}
-F^{0j}F_{jk}\te^{ki} -\fr{1}{2}\te^{jk}F_{kj}F_{0i}
\right) \approx 0 .  
&   \lb{scc2n2} 
\eeqa
$K_n^i$ and 
$L_n^i$ 
are some constants 
which should be determined by taking into account  
the other constraints when
they vanish strongly. 
The constraints (\ref{np1}) and
 (\ref{zo2}) are also second class. 
Structure of the constraints
is similar to the commuting case discussed in Section 2.
In fact, the number of physical phase space fields
is four.

In phase space, 
partition function 
associated with the parent action
for noncommutative $U(1)$ theory (\ref{npar}) is defined as
\be
\lb{tiz}
\ti{Z}=\int DP DP_D DF DA_D\ \ti{\De}
\exp\left\{i \int d^3x\left[P_D^{\mu}\dot{A}_{D\mu}+P_{\mu\nu}
\dot{F}^{\mu\nu}-H_{NPC}\right] \right\} .
\ee
Indices of the integration variables are suppressed.
We have adopted  the gauge fixing conditions 
\be
\ti{\La}^1=A_{D0}\approx 0, \qquad \ti{\La}^2=\del_iA_{Di} \approx 0,
\ee
for the first class constraints (\ref{nx1}) and (\ref{sla}).
Therefore, the measure $\ti{\De}$ is 
\be
\ti{\De}= 
\det \{\ti{\xi}^\al,\ti{\La}^\bt\}
{\rm det}^\fr{1}{2}\{\ti{\Phi}^a,
\ti{\Phi}^b\}
\prod_{\si =1}^2
\kd (\ti{\xi}^\si) \kd (\ti{\La}^\si) \prod_{c =1}^4 \kd(\ti{\Phi}^c).
\ee
Contribution of the first class constraints 
$\ti{\xi}^\al$ and their subsidiary conditions $\ti{\La}^\al $
to the measure is
\be
\det \{ \ti{\xi}^\al ,\ti{\La}^\bt \} = \det ( \del^2 ).
\ee
The second class constraints $\ti{\Phi}^a$
contribute to the measure as
\be
\lb{dscn}
{\rm det}^\fr{1}{2}\{\ti{\Phi}^a,\ti{\Phi}^b\}=
\det \left( \ep_{ijk}\del^iM_1^jM_2^k  \right)
\det \left( \ep_{ijk}\del^iC_1^jC_2^k  \right) 
\det \left( -1+ \fr{1}{2} \te^{ij}F_{ji}\right).
\ee
$ \ep_{ijk}\del^iM_1^jM_2^k$ and 
$ \ep_{ijk}\del^iC_1^jC_2^k$ denote multiplication of
three linear differential operators and as
usual, determinants of them 
are defined
as multiplication of the eigenvalues 
of the linear operators.
The last term in
(\ref{dscn}) is to be interpreted as multiplication of the
value of
$\left( -1+ \fr{1}{2} \te^{ij}F_{ji}\right)$
over all space--time. The determinants should be regularized,
however as we will show,  our results are independent of their
regularizations.

Performing  functional integrations
over $F^{\mu\nu}$ and  $P_{\mu\nu}$ in (\ref{tiz}) we obtain
\beqa
\ti{Z} &=& \int D\mbf{A}_D D\mbf{P}_D
\kd(\mbf{\del} \cdot \mbf{P}_D)\kd(\mbf{\del} \cdot \mbf{A}_D)
{\rm det} (\del^2) \nonumber \\
&& \exp{\Big \{} i \int d^3x {\Big [} P_{Di}\dot{A}_D^i-
\fr{1}{2g^2}P_{Di}P_D^i
- \fr{g^2}{4}F_{Dij}F_D^{ij} \nonumber \\
&& + \fr{1}{2g^4}\ti{\te}^{0i}P_{Di}P_D^2 + \ti{\te}^{0i}
F_{Dij}F_D^{jk}P_{Dk}
+ \fr{1}{4}\ti{\te}^{0i}P_{Di} F_D^2 {\Big  ] \Big \}}. \lb{ztnc}
\eeqa
The determinant (\ref{dscn}) is canceled\footnote{Obviously, to obtain 
(\ref{ztnc}) one does not need to separate $\ti{\chi}^4_i$
as (\ref{scc2n1})-- (\ref{scc2n2}).}
when we used $\kd (\ti{\Phi}^a)$
to express the 
``redundant" fields 
$F^{\mu\nu},\ P_{\mu\nu}$ 
in terms of the 
``physical" fields $A_D^i,\ P_D^i.$ Obviously, there are 
other solutions of (\ref{scc2n1}) and (\ref{scc2n2}) which would 
be useful to express another 
 set of fields in terms of the remaining ones. We take the solution
yielding the  partition function which we desire. 
We observe that in (\ref{ztnc}) the exponential term is the 
first order action of the dual  noncommutative $U(1)$ theory
whose hamiltonian is (\ref{nchd}).

Like the commuting case discussed in Section 2, when 
$\ti{\Phi}^3=0$ is used to write 
\[
F_{ij} =\del_iA_j-\del_j A_i,
\]
we demand that  the constraint
\[
\del_iA^i\approx 0
\]
should be fulfilled.
Moreover, when we change the variables
 $F_{ij}\to A_i$ 
we choose  the domains of integrations
in (\ref{tiz}) such that (\ref{nor}) is satisfied.
Equipped with these, we 
perform  integrations over
the fields $A_{D\mu},\ P_{D\mu},\ P_{\mu\nu}$ 
in (\ref{tiz}) which yield
\beqa
\ti{Z} & =&\det g^{-4} \int D\mbf{A} DF^{0i}
\kd(\mbf{\del} \cdot \mbf{A})  \det (\del^2) 
\det \left( -1+ \fr{1}{2} \te^{ij}F_{ji}\right)
\nonumber \\
&& \kd\left(
\del_i(F^{0i}-F_{ij}\te^{jk}F_{k0}
-F^{0j}F_{jk}\te^{ki} -\fr{1}{2}\te^{jk}F_{kj}F_{0i})\right)
\nonumber \\
&& \exp {\Big \{} i \int d^3x {\Big [} \fr{1}{g^2}\dot{A}^i(F^{0i}
-F_{ij}\te^{jk} F_{k0}-F^{0j}F_{jk}\te^{ki}-
\fr{1}{2}\te^{jk}F_{kj}F_{0i}) \nonumber \\
&& +\fr{1}{2g^2}F_{0i}F^{0i}-\fr{1}{4g^2}F_{ij}F^{ij}+
\fr{1}{g^2}F^{0i}F^{0j}F_{jk}\te^{ki}-
\fr{1}{4g^2}\te^{jk}F_{jk}F_{0i}F^{0i} \nonumber \\
&&+\fr{1}{8g^2}\te^{ij}F_{ij}F_{kl}F^{kl} {\Big ] \Big \}}. \lb{npf2}
\eeqa
We made use of the fact that employing 
$\kd(\ti{\Phi}^a)\kd(\mbf{\del} \cdot \mbf{P}_D)
\kd(\mbf{\del} \cdot \mbf{A}_D)$
to express $P_D^i,\  A^i_D$ in terms of
$F_{0i}$ and $A_i$ gives the following contribution to the measure
\[
\left[ \det g^4 \det (\del^2)
 \det \left( \ep_{ijk}\del^iC_1^jC_2^k  \right) 
\det \left( \ep_{ijk}\del^iM_1^jM_2^k  \right) \right]^{-1}.
\]
To deal with $P_i$ which are the canonical momenta of $A_i,$ 
let us adopt the change of variables,
\be
\lb{fi}
g^2 P^i=
F^{0i}-F_{ij}\te^{jk}F_{k0}-F^{0j}F_{jk}\te^{ki}
-\fr{1}{2}\te^{jk}F_{kj}F_{0i},
\ee
by inspecting the terms multiplying 
$\dot{A}^i.$  Thus, 
the partition function (\ref{npf2}) is written as
\beqa
\ti{Z}& =& \det g^{-4}\int D\mbf{A} D\mbf{P} 
\kd(\mbf{\del} \cdot \mbf{P})\kd(\mbf{\del} \cdot \mbf{A}) 
{\rm det}(\del^2) \nonumber \\
&& \exp {\Big \{ }i \int d^3x {\Big [}
\dot{A}^i P_i-\fr{g^2}{2}P_iP^i
-\fr{1}{4g^2}F_{ij}F^{ij}-g^2\te^{ij}P_i P^k F_{jk} \nonumber \\
&& +\fr{g^2}{4}\te^{ij}F_{ji}P^2+\fr{1}{8g^2}\te^{ij}
F_{ij}F_{kl}F^{kl}{\Big ] \Big \}}.  \lb{npf1}
\eeqa
In the exponential factor of (\ref{npf1}) we recognize
the hamiltonian of the noncommutative 
$U(1)$  theory (\ref{ncho}). 

It could be possible to show that the canonical
momenta $P_i$ are given as in (\ref{fi}) using  
the Dirac brackets:  
\beqa
\{ F_{0i}(x),P_{Dj}(y) \}_{\rm Dirac} &=& \{ F_{0i},P_{0k} \}
 \{ P_{0k},\ti{\Phi}_l^4 \}^{-1}
\{ \ti{\Phi}_l^4,P_{Dj} \} = g^2 \ep_{jkl}
[\kd_i^k+F^{km}\te_{mi} \nonumber \\
&& + F_{im}\te^{mk}
+\fr{1}{2}\kd_i^k\te^{mn}F_{nm}] \del_x^l\kd^3(x-y). \lb{ldb}
\eeqa
Vanishing of (\ref{np2}) and (\ref{zo2}) strongly,
permit us to write
the left hand side of (\ref{ldb})
equivalently  as 
\be
\{ F_{0i}(x),P_{Dj}(y)\}_{\rm Dirac}=-\ep_{jkl}\del_y^k 
\{ F_{0i}(x),A^l(y)\}_{\rm Dirac}. \lb{ldba}
\ee  
By comparing the right hand sides of (\ref{ldb})  and  (\ref{ldba})
we observe that they are compatible when
\be
F_{0i}=-g^2(P_i+F_{ij}\te^{jk}P_k+F^{jk}\te_{ki}P_j
-\fr{1}{2}F_{jk}\te^{kj}P_i).
\ee
Solving this equation for $P_i$ 
at the first order in $\theta_{ij}$ gives rise to (\ref{fi}).

We adopt the normalization consistent with Section 2, to write
 partition function
of the  noncommutative $U(1)$ theory in phase space
as
\beqa
Z_{NC}& =&\det g^{-2}\int D\mbf{A} D\mbf{P} 
\kd(\mbf{\del} \cdot \mbf{P})\kd(\mbf{\del} \cdot \mbf{A}) 
{\rm det}(\del^2) \nonumber \\
&& \exp {\Big \{ }i \int d^3x {\Big [}
\dot{A}^i P_i-\fr{g^2}{2}P_iP^i
-\fr{1}{4g^2}F_{ij}F^{ij}-g^2\te^{ij}P_i P^k F_{jk} \nonumber \\
&& +\fr{g^2}{4}\te^{ij}F_{ji}P^2+\fr{1}{8g^2}\te^{ij}
F_{ij}F_{kl}F^{kl}{\Big ] \Big \}}.  \lb{npfg}
\eeqa
Accordingly, the dual partition function
is given by
\beqa
Z_{NCD}&=& \det g^2 \int D\mbf{A} D\mbf{P}
\kd(\mbf{\del} \cdot \mbf{P})\kd(\mbf{\del} \cdot \mbf{A})
{\rm det} (\del^2) \nonumber \\
&& \exp{\Big \{} i \int d^3x {\Big [} P_{i}\dot{A}^i-
\fr{1}{2g^2}P_{i}P^i
- \fr{g^2}{4}F_{ij}F^{ij} \nonumber \\
&& + \fr{1}{2g^4}\ti{\te}^{0i}P_{i}P^2 + \ti{\te}^{0i}
F_{ij}F^{jk}P_{k}
+ \fr{1}{4}\ti{\te}^{0i}P_{i} F^2 {\Big  ] \Big \}}, \lb{ztnk}
\eeqa
where we renamed $A_{Di},\ P_{Di}$ as $A_i,\ P_i .$

We conclude that in phase space, partition functions
for the  noncommutative $U(1)$ theory
and its dual are the same
\be
\lb{res}
Z_{NC}=Z_{NCD}.
\ee 
This result
demonstrates that strong--weak duality 
transformation 
is  helpful 
to make calculations
in weak coupling regions to 
extract
information about 
 physical quantities 
in the strong  coupling regions.

We would like to emphasize the difference 
between  the results
obtained for the commutative  case (\ref{peq})
and  for the noncommutative $U(1)$ theory  (\ref{res}). 
In $U(1)$ gauge theory, partition functions for the initial 
and the dual theories, (\ref{ytmh}) and (\ref{ythd}), 
are equivalent and they 
are related 
with the map  $g \to g^{-1}.$ However, the partition function 
of  noncommutative $U(1)$ (\ref{npfg}) does not yield the 
partition function of its dual (\ref{ztnk}) by only inverting the coupling 
constant, although they are equivalent.

Application of the approach presented here to 
noncommutative
supersymmetric $U(1)$ gauge theory 
whose parent actions were studied  in \cite{duy},
may shed light on the duality symmetry of 
the supersymmetric noncommutative theory.

We dealt with free theories, although
introducing source terms into
the starting  path integral (\ref{tiz}) 
to gain insight about relations of the  Green functions of 
the noncommutative $U(1)$ theory and its dual
is very important.


\newpage

\begin{thebibliography}{99}

\bibitem{bus}T.H. Buscher, 
{\it A symmetry of the string background field equations,
Phys. Lett.} $\mbf{B}$ $\mbf{194}$ (1987) 59;
{\it Path-integral derivation of
quantum duality in nonlinear sigma models, Phys. Lett. }
$\mbf{B}$ $\mbf{201}$ (1988) 466.

\bibitem{loz}Y. Lozano, {\it S-duality in gauge theories
as a canonical transformation, Phys. Lett.} $\mbf{B}$ $\mbf{364}$
(1995) 19 [hep-th/9508021].

\bibitem{sw}N. Seiberg and E. Witten, {\it String theory
and non-commutative geometry, J. High Energy Phys.} {\bf 09} (1999)
032 [hep-th/9908142].

\bibitem{grs} O.J. Ganor, G. Rajesh and S. Sethi,
{\it Duality and non-commutative gauge theory,
Phys. Rev. }$\mbf{D}$ $\mbf{62}$ (2000) 125008 [hep-th/0005046].

\bibitem{dy}\"{O}.F. Dayi  and  B. Yap\i\c{s}kan,
{\it Hamiltonian formulation of non-commutative D3-brane,
J. High Energy Phys.} {\bf 10} (2002) 022 [hep-th/0208043]. 

\bibitem{abt}Y. Abe, R. Banerjee and I. Tsutsui,
{\it Duality symmetry and plane waves in non-commutative
electrodynamics, Phys. Lett.} $\mbf{B}$ $\mbf{573}$ (2003) 248
[hep-th/0306272].

\bibitem{fra}E. S. Fradkin, {\it in New Developments in Relativistic 
Quantum Field Theory, Proceedings of the Xth Winter School
of Theoretical Physics, Karpacz, Poland,} 1973 [{\it Acta Univ.
Wratislav.} No. 207, Poland (1973)].

\bibitem{sen}P. Senjanovic, {\it Path integral quantization 
of field theories with second class constraints,
Annals of Physics} {\bf 100} (1976) 227.

\bibitem{kru}S.I. Kruglov, {\it Maxwell's Theory on Non-Commutative 
Spaces and Quaternions, Annales Fond.Broglie} {\bf 27} (2002) 343
[hep-th/0110059].

\bibitem{duy}\"{O}.F. Dayi, K. \"{U}lker  and  B. Yap\i\c{s}kan,
{\it Duals of noncommutative supersymmetric $U(1)$ gauge
theory, 
J. High Energy Phys.} {\bf 10} (2003) 010 [hep-th/0309073]. 


\end{thebibliography}
\end{document}